\documentclass[aps,prl,preprint,superscriptaddress,nofootinbib]{revtex4-1}

\usepackage{lineno}

\usepackage{color,soul}

\usepackage{amsmath}

\usepackage{textcomp}

\usepackage{graphicx}

\begin{document}

%\preprint{}

\title{Measurement of Cosmic-ray Electrons at TeV Energies by VERITAS}


%\author{D. Staszak et al.}
%\email[]{staszak@kicp.uchicago.edu}
\author{A.~Archer}
%\affiliation{Department of Physics, Washington University, St. Louis, MO 63130, USA}
\affiliation{Department of Physics and Astronomy, Purdue University, West Lafayette, IN 47907, USA}
\author{W.~Benbow}
\affiliation{Fred Lawrence Whipple Observatory, Harvard-Smithsonian Center for Astrophysics, Amado, AZ 85645, USA}
\author{R.~Bird}
\affiliation{Department of Physics and Astronomy, University of California, Los Angeles, CA 90095, USA}
\author{R.~Brose}
\affiliation{Institute of Physics and Astronomy, University of Potsdam, 14476 Potsdam-Golm, Germany}
\affiliation{DESY, Platanenallee 6, 15738 Zeuthen, Germany}
\author{M.~Buchovecky}
\affiliation{Department of Physics and Astronomy, University of California, Los Angeles, CA 90095, USA}
\author{J.~H.~Buckley}
\affiliation{Department of Physics, Washington University, St. Louis, MO 63130, USA}
\author{V.~Bugaev}
\affiliation{Department of Physics, Washington University, St. Louis, MO 63130, USA}
\author{M.~P.~Connolly}
\affiliation{School of Physics, National University of Ireland Galway, University Road, Galway, Ireland}
\author{W.~Cui}
\affiliation{Department of Physics and Astronomy, Purdue University, West Lafayette, IN 47907, USA}
\affiliation{Department of Physics and Center for Astrophysics, Tsinghua University, Beijing 100084, China.}
\author{M.~K.~Daniel}
\affiliation{Fred Lawrence Whipple Observatory, Harvard-Smithsonian Center for Astrophysics, Amado, AZ 85645, USA}
\author{Q.~Feng}
\affiliation{Physics Department, McGill University, Montreal, QC H3A 2T8, Canada}
\author{J.~P.~Finley}
\affiliation{Department of Physics and Astronomy, Purdue University, West Lafayette, IN 47907, USA}
\author{L.~Fortson}
\affiliation{School of Physics and Astronomy, University of Minnesota, Minneapolis, MN 55455, USA}
\author{A.~Furniss}
\affiliation{Department of Physics, California State University - East Bay, Hayward, CA 94542, USA}
\author{G.~Gillanders}
\affiliation{School of Physics, National University of Ireland Galway, University Road, Galway, Ireland}
\author{M.~H\"utten}
\affiliation{DESY, Platanenallee 6, 15738 Zeuthen, Germany}
\author{D.~Hanna}
\affiliation{Physics Department, McGill University, Montreal, QC H3A 2T8, Canada}
\author{O.~Hervet}
\affiliation{Santa Cruz Institute for Particle Physics and Department of Physics, University of California, Santa Cruz, CA 95064, USA}
\author{J.~Holder}
\affiliation{Department of Physics and Astronomy and the Bartol Research Institute, University of Delaware, Newark, DE 19716, USA}
\author{G.~Hughes}
\affiliation{Fred Lawrence Whipple Observatory, Harvard-Smithsonian Center for Astrophysics, Amado, AZ 85645, USA}
\author{T.~B.~Humensky}
\affiliation{Physics Department, Columbia University, New York, NY 10027, USA}
\author{C.~A.~Johnson}
\affiliation{Santa Cruz Institute for Particle Physics and Department of Physics, University of California, Santa Cruz, CA 95064, USA}
\author{P.~Kaaret}
\affiliation{Department of Physics and Astronomy, University of Iowa, Van Allen Hall, Iowa City, IA 52242, USA}
\author{P.~Kar}
\affiliation{Department of Physics and Astronomy, University of Utah, Salt Lake City, UT 84112, USA}
\author{N.~Kelley-Hoskins}
\affiliation{DESY, Platanenallee 6, 15738 Zeuthen, Germany}
\author{M.~Kertzman}
\affiliation{Department of Physics and Astronomy, DePauw University, Greencastle, IN 46135-0037, USA}
\author{D.~Kieda}
\affiliation{Department of Physics and Astronomy, University of Utah, Salt Lake City, UT 84112, USA}
\author{M.~Krause}
\affiliation{DESY, Platanenallee 6, 15738 Zeuthen, Germany}
\author{F.~Krennrich}
\affiliation{Department of Physics and Astronomy, Iowa State University, Ames, IA 50011, USA}
\author{S.~Kumar}
\affiliation{Department of Physics and Astronomy and the Bartol Research Institute, University of Delaware, Newark, DE 19716, USA}
\author{M.~J.~Lang}
\affiliation{School of Physics, National University of Ireland Galway, University Road, Galway, Ireland}
\author{T.~T.Y.~Lin}
\affiliation{Physics Department, McGill University, Montreal, QC H3A 2T8, Canada}
\author{G.~Maier}
\affiliation{DESY, Platanenallee 6, 15738 Zeuthen, Germany}
\author{S.~McArthur}
\affiliation{Department of Physics and Astronomy, Purdue University, West Lafayette, IN 47907, USA}
\author{P.~Moriarty}
\affiliation{School of Physics, National University of Ireland Galway, University Road, Galway, Ireland}
\author{R.~Mukherjee}
\affiliation{Department of Physics and Astronomy, Barnard College, Columbia University, NY 10027, USA}
\author{S.~O'Brien}
\affiliation{School of Physics, University College Dublin, Belfield, Dublin 4, Ireland}
\author{R.~A.~Ong}
\affiliation{Department of Physics and Astronomy, University of California, Los Angeles, CA 90095, USA}
\author{A.~N.~Otte}
\affiliation{School of Physics and Center for Relativistic Astrophysics, Georgia Institute of Technology, 837 State Street NW, Atlanta, GA 30332-0430}
\author{A.~Petrashyk}
\affiliation{Physics Department, Columbia University, New York, NY 10027, USA}
\author{M.~Pohl}
\affiliation{Institute of Physics and Astronomy, University of Potsdam, 14476 Potsdam-Golm, Germany}
\affiliation{DESY, Platanenallee 6, 15738 Zeuthen, Germany}
\author{E.~Pueschel}
\affiliation{DESY, Platanenallee 6, 15738 Zeuthen, Germany}
\author{J.~Quinn}
\affiliation{School of Physics, University College Dublin, Belfield, Dublin 4, Ireland}
\author{K.~Ragan}
\affiliation{Physics Department, McGill University, Montreal, QC H3A 2T8, Canada}
\author{P.~T.~Reynolds}
\affiliation{Department of Physical Sciences, Cork Institute of Technology, Bishopstown, Cork, Ireland}
\author{G.~T.~Richards}
\affiliation{School of Physics and Center for Relativistic Astrophysics, Georgia Institute of Technology, 837 State Street NW, Atlanta, GA 30332-0430}
\author{E.~Roache}
\affiliation{Fred Lawrence Whipple Observatory, Harvard-Smithsonian Center for Astrophysics, Amado, AZ 85645, USA}
\author{C.~Rulten}
\affiliation{School of Physics and Astronomy, University of Minnesota, Minneapolis, MN 55455, USA}
\author{I.~Sadeh}
\affiliation{DESY, Platanenallee 6, 15738 Zeuthen, Germany}
\author{M.~Santander}
\affiliation{Department of Physics and Astronomy, University of Alabama, Tuscaloosa, AL 35487, USA}
\author{G.~H.~Sembroski}
\affiliation{Department of Physics and Astronomy, Purdue University, West Lafayette, IN 47907, USA}
\author{D.~Staszak}
\email{d.staszak@gmail.com}
\affiliation{Enrico Fermi Institute, University of Chicago, Chicago, IL 60637, USA}
\author{I.~Sushch}
\affiliation{DESY, Platanenallee 6, 15738 Zeuthen, Germany}
\author{S.~P.~Wakely}
\affiliation{Enrico Fermi Institute, University of Chicago, Chicago, IL 60637, USA}
\author{R.~M.~Wells}
\affiliation{Department of Physics and Astronomy, Iowa State University, Ames, IA 50011, USA}
\author{P.~Wilcox}
\affiliation{Department of Physics and Astronomy, University of Iowa, Van Allen Hall, Iowa City, IA 52242, USA}
\author{A.~Wilhelm}
\affiliation{Institute of Physics and Astronomy, University of Potsdam, 14476 Potsdam-Golm, Germany}
\affiliation{DESY, Platanenallee 6, 15738 Zeuthen, Germany}
\author{D.~A.~Williams}
\affiliation{Santa Cruz Institute for Particle Physics and Department of Physics, University of California, Santa Cruz, CA 95064, USA}
\author{T.~J~Williamson}
\affiliation{Department of Physics and Astronomy and the Bartol Research Institute, University of Delaware, Newark, DE 19716, USA}
\author{B.~Zitzer}
\email{bzitzer@physics.mcgill.ca}
\affiliation{Physics Department, McGill University, Montreal, QC H3A 2T8, Canada}

\collaboration{The VERITAS Collaboration}
\noaffiliation
%\thanks{}


\date{\today}

\begin{abstract}

Cosmic-ray electrons and positrons (CREs) at GeV-TeV energies are a unique 
probe of our local Galactic neighborhood. 
CREs lose energy rapidly via synchrotron radiation and inverse-Compton 
scattering processes while propagating within the Galaxy and these losses 
limit their propagation distance.
For electrons with TeV energies, the limit is on the order of a kiloparsec.
Within that distance there are only a few known astrophysical objects capable of
accelerating electrons to such high energies. 
It is also possible that the CREs are the products of the annihilation or decay 
of heavy dark matter (DM) particles.
VERITAS, an array of imaging air Cherenkov telescopes 
in southern Arizona, USA, is primarily utilized for gamma-ray astronomy, 
but also simultaneously collects CREs during all observations. 
We describe our methods of identifying CREs in VERITAS data
and present an energy spectrum, extending from 300 GeV  to 5 TeV, obtained
from approximately 300 hours of observations. A single power-law fit is ruled out in VERITAS data. We find that the spectrum of CREs is
consistent with a broken power law, with a break energy at 
710 $\pm$ 40$_{stat}$ $\pm$ 140$_{syst}$ GeV.

\end{abstract}

%\keywords{}

\maketitle

%\section{}
\section{Introduction\label{SecIntro}}

Despite constituting only a small fraction of the total cosmic-ray flux, 
cosmic-ray electrons and positrons (CREs) provide an important and unique 
probe of our local Galactic neighborhood. 
They rapidly lose energy while propagating in the Galaxy via synchrotron 
radiation and inverse-Compton scattering processes.
This limits the propagation distance for TeV electrons to of 
order $\sim$1 kpc~\cite{COWSIK}~\cite{Grasso2009}~\cite{Pohl1998}
and implies that CREs at TeV energies can provide constraints on local 
cosmic-ray accelerators and diffusion effects.

The ${\it Fermi}$-LAT~\cite{LAT} collaboration and the AMS-02 
collaboration~\cite{AMS}, have measured the CRE spectrum 
up to energies of $\sim$1 TeV. More recently, both the 
DAMPE~\cite{DAMPE} and CALET~\cite{CALET} collaborations have measured the 
CRE spectrum to a few TeV with excellent energy resolution.
At higher energies these instruments run out of statistics 
due to the combination of the steep CRE spectrum and their relatively 
small acceptances.
Ground-based imaging atmospheric Cherenkov telescopes (IACTs)
can extend the CRE spectrum to higher energies due to their large 
collection areas ($\sim10^{5}$ m$^{2}$).
H.E.S.S.~\cite{HESS1, HESS2} and MAGIC~\cite{MAGIC} have demonstrated 
this ability and have measured the CRE spectrum up to energies of several TeV.
Their results agree 
with the space-based measurements, within systematic uncertainties,
in the energy range where the sensitivities overlap.
The combined picture that has emerged is one where the CRE spectrum can be 
described by a simple power law from $\sim$10 GeV up to just below $\sim$1 TeV. 
At higher energy H.E.S.S. sees a spectral steepening\footnotemark[1]  while 
MAGIC data are consistent with a single power law up to 
$\sim$3 TeV, although with larger statistical uncertainties. The DAMPE and CALET data also see a break in the CRE spectrum at $\sim$1 TeV~\cite{CALET}~\cite{DAMPE}. 
\footnotetext[1]{The H.E.S.S. collaboration has recently reported preliminary results, obtained with higher statistics over an extended energy range, which support this trend.}

The inclusive CRE spectrum is understood to have contributions from both 
electrons and positrons and some instruments are able to separate the two
components.
The energy dependence of the positron fraction, $e^+/(e^+ + e^-)$, has been measured 
for energies greater than 10 GeV by the HEAT~\cite{HEAT}, PAMELA~\cite{PAM}, 
{\it Fermi}-LAT~\cite{LATfrac}, and AMS-02~\cite{AMSfrac} collaborations.
The fraction is found to rise with increasing energy up to $\sim$200 GeV, 
above which it appears to flatten out.
Positrons are believed to be produced mainly in interactions 
between cosmic rays and interstellar gas and this results in a positron
fraction that decreases with energy. 
%Positrons cool to about an MeV before annihilation so depletion is unlikely to affect 
%the positron fraction at higher energies. 
The unexpected increase could imply the existence of additional local 
sources such as pulsars or supernovae remnants.
A more exotic explanation for the excess would be the annihilation or decay 
of DM particles. 
More conventional methods such as an improved propagation model~\cite{prop} 
or a better accounting of secondary production might explain the results.
A full understanding of this situation will require detailed input about both 
the positron fraction and the CRE spectrum.
Given that the CRE spectrum, including its behavior at TeV energies, is such 
an integral component, it is important for all
instruments capable of making such measurements to contribute.

\section{Methods\label{SecMeth}}
%\subsubsection{}

VERITAS is an IACT instrument. 
In contrast to space-borne detectors like {\it Fermi}-LAT and AMS-02,
IACTs do not measure the primary particles (electrons, photons, nuclei) directly.
They detect the Cherenkov light generated by charged particles in 
air showers that result from the impact of the primaries on the upper atmosphere.
As such, there is no information on the charge of the primary; electrons
and positrons cannot be measured separately.
Therefore, CREs in this work refer to the sum of positrons and electrons.
Air showers caused by gamma rays and electrons are electromagnetic and are 
systematically different from the hadronic showers caused by cosmic-ray nuclei.
Indeed, these differences are often exploited to separate weak gamma-ray signals from 
large hadronic backgrounds.
However, showers due to electrons are essentially identical to those from 
gamma rays and there is no practical way to discriminate between these two types of primary particles.

In the following we will use gamma-hadron separation techniques to isolate 
a gamma-like signal, which will include the electron signal but also a 
background of gamma rays.
To limit the background contribution we will exclude data from regions around known VHE gamma-ray sources.
The Galactic plane, a known source of diffuse gamma 
radiation, is also excluded.
There is a limited amount of extra-Galactic diffuse gamma radiation but recent 
measurements by {\it Fermi}-LAT~\cite{diffuseLAT} up to $\sim$800 GeV show that
it is of the same order of magnitude as that from discrete sources (mostly 
blazars) and is described by a power law, with index of about 2.3, and an 
exponential cut-off at 250 GeV. Based on this {\it Fermi}-LAT result, we estimate 
perhaps one out of a thousand CRE-like events is actually a diffuse gamma ray at 250 
GeV. This ratio falls off rapidly at higher energies. Contamination from the Fermi 
bubbles \cite{FermiBubble} is also possible and we did not explicitly exclude data in 
those regions. The spectrum of the Fermi bubbles show an exponential cutoff at 
$\sim$110 GeV, so it is unlikely to affect the results in this work.
We will therefore assume that the gamma-ray contamination in our 
CRE candidate events is negligible.  

Separation between hadronic showers and gamma or electron-like showers is not fully efficient. 
After the application of data selection criteria (cuts) 
a substantial hadronic background remains, often larger than the signal itself.
When measuring gamma-ray sources this background can be determined and 
subtracted using one of the techniques described in Berge et. al. 2007~\cite{Berge}.
These techniques rely on the presence of a source, known or postulated, 
in the FOV, surrounded by a source-free region from which to estimate the background.
For CREs this is not the case;
the intensity is essentially isotropic so there is no way to experimentally 
estimate the background and one must rely on simulation-based calculations.
This introduces additional systematic uncertainties into the CRE estimate.

\subsection{VERITAS}

Data presented here were collected with the VERITAS array, located 
at the Fred Lawrence Whipple Observatory (FLWO) in southern Arizona, USA 
(31$^{\circ}$ 40$'$ N, 110$^{\circ}$ 57$'$ W,  1.3 km a.s.l.).
The array comprises four identical telescopes~\cite{Holder}, 
each with a 12-m Davies-Cotton reflector 
that collects Cherenkov light and directs it onto a camera made up of 499 
photomultiplier tubes (PMTs). The telescope FOV has a diameter of 3.5$^\circ$.

Outputs from the PMTs are continuously digitized by 500 Msps 8-bit flash analog-to-
digital converters (FADCs) and written to a buffer. A three-level trigger, requiring 
a three-fold coincidence of neighboring pixels, each of which 
passes a discriminator threshold in at least two 
telescopes, initiates readout of the digitized signals.

VERITAS has been running with four telescopes since 2007.
During the summer of 2009 we repositioned one of the telescopes to improve the
array sensitivity and in summer 2012 we installed PMTs with higher 
quantum efficiency to lower the energy threshold~\cite{Kieda}. 
For this study we use data acquired between September 2009 and July 2012, with a 
narrow zenith angle range. The choice of this period was motivated by the fact 
that at the time this study started, this was the period of the largest and best-understood data set available.
During this period VERITAS had a gamma-ray energy threshold of approximately 100 
GeV, an energy resolution of 15-20\% between 0.1 and 10 TeV, 
and an angular resolution (68\% containment) of less than 0.1$^\circ$ at 
1 TeV~\cite{Park}. CRE events have identical resolutions to gamma rays.  

\subsection{Data Selection}

Since the CRE flux is isotropic we can use data from all VERITAS 
observations in our study, subject to the following selection criteria: 

\begin{itemize}

\item{clear weather,}

\item{all telescopes operational,}

\item{zenith angle between 15$^\circ$ and 25$^\circ$,}

\item{observation field at least 10$^\circ$ away from the Galactic plane.}

\end{itemize}

These cuts resulted in a data set constituting 296 hours of live time. The range 
of zenith angles between 15 and 25 degrees is the window containing the most data 
available. Additionally, the best sensitivity of VERITAS is for zenith angles less 
than 25 degrees \cite{Park}. The restricted zenith angle range and single 
detector configuration corresponds to where most VERITAS observations were made at the 
time this study started. Using the restricted range greatly reduced the number of 
simulations required for this study.
The data were processed using standard VERITAS reconstruction software and the 
resulting events were subjected to further cuts:

\begin{itemize} 

\item{Every telescope should have a good shower image according to the standard 
VERITAS reconstruction package~\cite{evdisp}. This reduces the overall systematic 
uncertainties in both the energy and position reconstruction of the shower. Above 
1 TeV, 95\% of simulated CRE events have good shower images in all four telescopes.}

\item{The direction of the reconstructed shower should point to within a degree 
of the nominal array pointing direction. VERITAS has peak sensitivity using events within one degree of the pointing direction \cite{Park}.}

\item{The shower axis should intersect the array plane within 200 m of the array 
center. Roughly 90\% of simulated CRE events reconstruct at distances less than this, therefore this cut only removes a small number of potential signal events from the analysis.}

\item{For observation fields that contain known or potential gamma-ray sources,
the shower axis should not point within 0.187$^{\circ}$ of those 
objects. (The actual cut is on the square of this value, 0.035 deg$^{2}$.) This is wider than the gamma-ray PSF and should remove any 
significant gamma-ray contamination in the CRE data. The list of known gamma-ray 
sources was obtained from TeVCat \cite{TeVCat}. 
Potential gamma-ray sources that were the \textit{a priori} targets of VERITAS 
observations were also excluded.}

\end{itemize}

Some of these cuts are more restrictive than in most VERITAS gamma-ray analyses, 
resulting in improved data/Monte Carlo agreement with the proton simulations. 
The cuts were optimized to improve the agreement of
diffuse proton simulations with proton data at both the
single-telescope image and array-wide levels, and to improve the agreement of gamma-ray
simulations with Crab Nebula data. The goal of applying these additional cuts was to 
reduce the overall systematic uncertainty in the CRE measurement. 

\subsection{Monte Carlo Simulations}

As explained previously, extracting a CRE signal requires extensive
use of Monte Carlo (MC) simulations. Showers initiated by electrons, protons, and helium 
nuclei were generated, with arrival directions up to 4$^\circ$ away from the array 
pointing direction.
This range was to cover the VERITAS FOV and it accommodates any edge effects; 
adding showers from larger angles had no effect. Each particle species was generated with an energy distribution of $dN/dE \sim E^{\Gamma}$ with $\Gamma$ = \textminus2 to increase statistics at higher energies. The proton energy distribution was then reweighed to $\Gamma$ = \textminus2.7 to agree with experimental data over the range we are sensitive to~\cite{Maurin}.
 
The CORSIKA 6.970 package~\cite{corsika} was used, 
with the QGSJetII.3~\cite{qgsjet} and
URQMD 1.3cr~\cite{urqmd1, urqmd2} event generators, to produce files of ground-level 
Cherenkov photons. Newer versions of CORSIKA have significantly improved 
simulations of hadronic interactions at the highest energies, but the energy range 
here is relatively unaffected.
These were processed with the GrISUDet 5.0.0~\cite{grisu} VERITAS detector 
modeling program before analysis with the one of the standard VERITAS event-reconstruction packages (EventDisplay~\cite{evdisp}).
A smaller set of proton-initiated showers was generated using the 
SIBYLL package~\cite{sibyll} and used to test for consistency. 
Results agreed within systematic uncertainties with those from QGSJetII.3.

\subsection{Boosted Decision Trees}

Cherenkov light from an electromagnetic (e.g. gamma-initiated) shower produces 
an approximately elliptical image in the camera plane of an IACT.
Images from hadronic (proton-initiated) showers are less regular in shape and 
can have large fluctuations in their morphology.
This forms the basis for all schemes of background rejection in IACT-based 
gamma-ray astronomy~\cite{fegan}. 
Traditionally, one characterizes the images using a set of parameters suggested
by Hillas~\cite{hillas}. Two of these are the length and width of the ellipse.
Images from hadronic showers are typically longer and wider than those
from electromagnetic showers.

The standard VERITAS analysis packages make use of image parameters and rely
on cuts placed separately on each of them to extract signals; these are 
called ``box cuts".
In this work we make use of boosted decision trees (BDTs)
to reject (hadronic) background and retain (electromagnetic) signal.
As explained in~\cite{BDT}, BDTs are a multivariate analysis technique.
They combine several input variables in such a way as to produce a single output
variable which describes, in this application, how ``electromagnetic-like" 
a shower is. 
The BDTs are trained with a combination of data and simulations and
provide the ability to utilize non-linear correlations between
training variables when classifying data.
They have been shown~\cite{BDT} 
to improve VERITAS sensitivity for a variety of 
gamma-ray source types compared with the traditional box-cut analyses.
Apart from the improved sensitivity offered by BDTs, they are useful because
their output is  
a single number that can be used in likelihood fitting procedures, as shown later
in this paper. These methods are essentially the same as those used by the H.E.S.S. 
and MAGIC collaborations~\cite{HESS1,HESS2,MAGIC} for CRE spectra, except for utilizing BDTs instead of Random Forests for hadron discrimination.

The BDTs used in this work were constructed using 
the Toolkit for Multivariate Data
Analysis (TMVA) component~\cite{TMVA} of the CERN ROOT package~\cite{root}. 
We used four array-level variables as input to the BDTs:

\begin{itemize}

\item{mean reduced scaled width (MRSW),} \cite{MSRW}

\item{mean reduced scaled length (MRSL),} \cite{MSRW}

\item{emission height, and}

\item{$\chi^{2}(E)$, which is defined as the RMS of the energy estimates from the individual telescopes. The mean of the individual telescope energy estimates is accepted as the reconstructed shower energy.}

\end{itemize}

These variables are described in more detail in~\cite{BDT}. 
The first two are related to the lengths and widths of the camera images. 
The third variable is the estimate of the height of maximum development of the
shower, made using data from all telescopes. On average, electromagnetic showers have a 
slightly lower value for this than do hadronic showers.
The fourth quantifies how well the different telescopes agree on the energy 
of the shower. $\chi^{2}(E)$ tends to be larger for hadronic showers.

The BDTs were trained with a signal data set of 6.4 million simulated 
diffuse electron events and a background data set consisting of 5.7 million events 
chosen randomly from the full data set. Both the signal and training samples pass the cuts described in the previous section.
Note that the ``background" events, chosen in this way, contain
an admixture of ``signal" events and this compromises the training 
somewhat. 
However, since CREs make up a small fraction of the total cosmic-ray flux,
this is a higher-order effect.
The use of data rather than simulated hadronic showers requires fewer CPU hours and is less dependent on the fidelity of the hadronic simulations.

Independent signal and background data sets were used to check for over-training
of the BDTs; no evidence for this was found.
The BDTs were also tested by using them for gamma-hadron separation in an 
analysis of Crab Nebula data.
The reconstructed energy spectrum was consistent with previous measurements.

\begin{figure}
\begin{center}
\includegraphics[width = 5.0in]{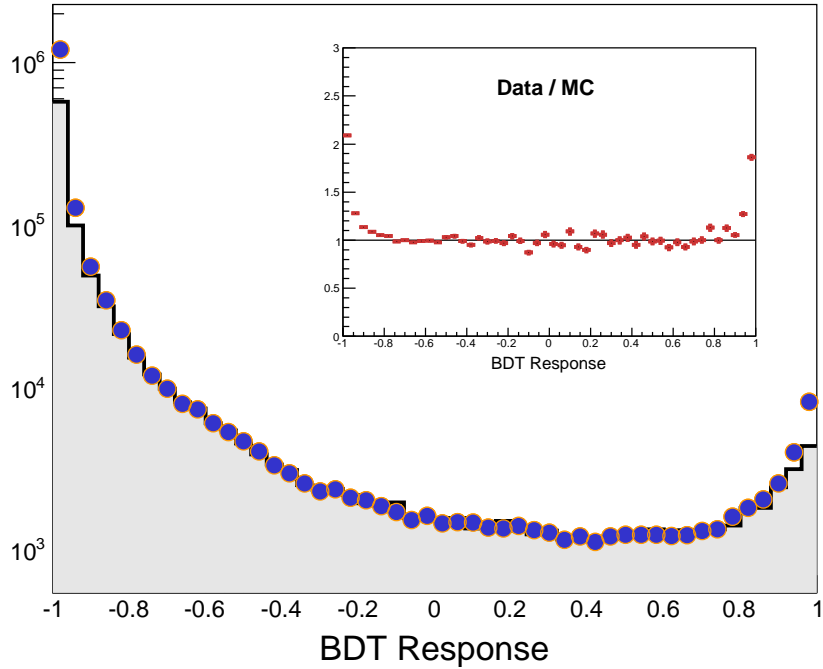}
\caption{
CRE BDT response parameter for the full data set (symbols) 
and for simulated proton-induced showers (histogram) in the energy range between 630 GeV and 1 TeV. 
The plot in the inset is the ratio of the two.
Hadronic showers are assigned BDT values close to -1.0 while electromagnetic
showers are assigned values near 1.0.
}
\label{offPlot}
\end{center}
\end{figure}

After training, the BDTs were used in classifying events from the data set
used for this study.
Each shower was assigned a BDT response value, from \textminus1.0 to 1.0, indicating
whether the shower was hadronic (\textminus1.0) or electromagnetic (1.0). 
Figure \ref{offPlot} shows a plot of BDT responses for the full 
data set compared with those from a set of simulated proton showers.
The agreement is very good except near the limits of the distribution.
There are excesses of data over proton simulations near \textminus1.0 and near 1.0.
At \textminus1.0 we expect an excess from helium and higher-Z primaries, particularly 
since helium makes up $\sim20\%$ of the overall cosmic-ray flux~\cite{pdg}.
We investigated the BDT response for a set of simulated
helium events and found it to 
peak at \textminus1.0 and fall off faster than that for simulated protons, 
supporting our interpretation of the excess.
There is a second excess at the opposite end of the response histogram.
This is the CRE signal, which is present in the observational data but not the hadronic
background simulations.

In the following, we select events with BDT response values greater than 0.7
and employ a binned extended likelihood fitting method 
within this region to extract the contributions of electron and proton events 
to the total.
Using the shape of the BDT response distribution from electron and proton simulations, we 
find the relative contributions that best describe the data.
We do not include contributions from helium and higher-Z shower events since 
they are sufficiently rejected in this region by the BDTs.

\section{Results\label{SecRes}}

\begin{figure}
\begin{center}
\includegraphics[width = 6.0in]{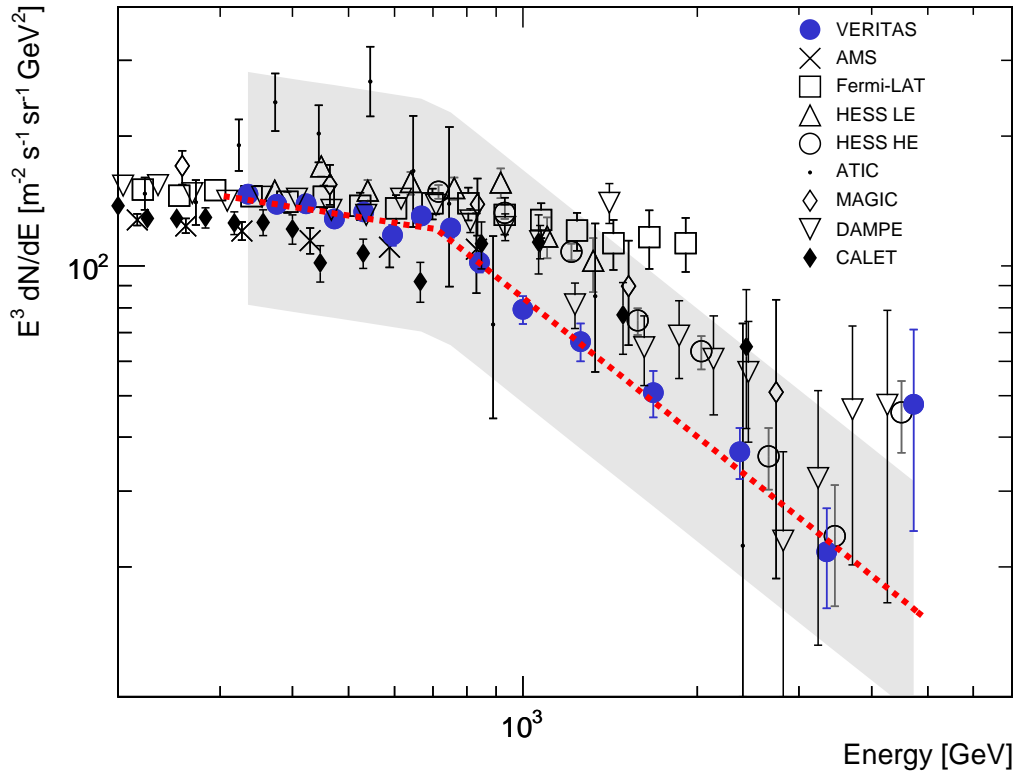}

\caption{
Spectrum of CREs between 300 GeV and 5 TeV, 
as measured by VERITAS along with previously published measurements.
Error bars are statistical; systematic uncertainties are indicated by the 
gray band.
%this needs up-to-date numbers from AMS-02 (and Fermi?) and a reference for ATIC - also need to get rid of the ICRC 2015 label
}
\label{MONEY}
\end{center}
\end{figure}

We show in Figure \ref{MONEY} the VERITAS CRE energy spectrum 
between 300 GeV and 5 TeV. These results are also summarized in Table 1.
The spectrum steepens at higher energies and is best described by 
two power laws, with a break between them as one index transitions to another.
The best fit for this break energy is found to be (710 $\pm$ 40$_{stat}$ $\pm$ 140$_{syst}$ ) GeV, 
with best-fit spectral indices below (above) this energy of $-$3.2 
$\pm$ 0.1$_{stat}$ ($-$4.1 $\pm$ 0.1$_{stat}$).
The chi-squared per degree-of-freedom ($\chi^{2}/dof$) of this fit is 9.71/11. 
Analytically, this fit is described by:

\begin{equation}
	F(E)=\begin{cases}k(E/E_{b})^{\Gamma_{1}} &\mbox{if } E \leq E_{b}\\
    	k(E/E_{b})^{\Gamma_{2}} &\mbox{if } E  > E_{b}\end{cases},
\end{equation}

where $E_{b}$ is the break energy, $k$ is the flux value at the break energy and $\Gamma_{1}  ~(\Gamma_{2})$ is the index below (above) the break energy.

In addition to a broken power law, a single power-law fit was performed, giving a 
significantly worse fit to the data, with a $\chi^{2}/dof$ of 76.5/13. A power law with 
an exponential cutoff was also fit to the data, yielding a $\chi^{2}/dof$ of 26.0/12.

The gray band represents the systematic uncertainty, which is dominated 
by the $\sim20\%$ uncertainty on the VERITAS absolute energy scale.
This translates into a +64\%/$-$33\% (+98\%/$-$43\%) systematic uncertainty 
for a spectral index of $-$3.2 ($-$4.1).

Additional tests on the robustness of the analysis were performed:

\begin{itemize}

\item{the assumed spectral index for the energies of the simulated protons was varied 
by $\pm$10\% from its nominal value,}

\item{BDT cut values 0.8 and 0.9  were investigated and}

\item{Hadronic-shower rejection was performed using cuts on the MRSW 
distributions and not the BDT response, 
effectively testing the entire machinery but not using BDTs.}

\end{itemize}

All changes resulted in final data points consistent with the quoted 
systematic uncertainties. It should be noted that the proton spectrum is well-measured in this energy range, so 10\% is considered very conservative.
For the non-BDT analysis a loss of efficiency occurred, resulting in larger
statistical uncertainties, and energy bins above 2 TeV could not be recovered.

\begin{table}[ht]
\begin{center}
\begin{tabular}{ p{2.5cm} | p{3.5cm} | p{2cm} | p{3cm} | p{5cm} }
\hline\hline
	Energy $[$TeV$]$ & Energy Bin $[$TeV$]$ & $N_{events}$ & Electron Fraction & Flux
    $[$cm$^{-2}$s$^{-1}$TeV$^{-1}]$ \\

\hline

  0.335  &  0.316  -  0.355  &	27529 & 0.152 $\pm$ 0.005 & 
  3.140$\times10^{-10}$ $\pm$  9.790$\times10^{-12}$ \\
  0.376  &  0.355  -  0.398  &  21330 &	0.159 $\pm$ 0.006 & 
  2.104$\times10^{-10}$ $\pm$  8.469$\times10^{-12}$\\
  0.422  &  0.398  -  0.447  &  15704 &	0.187 $\pm$ 0.007 & 
  1.491$\times10^{-10}$ $\pm$  5.681$\times10^{-12}$\\
  0.473  &  0.447  -  0.501  &  14787 &	0.154 $\pm$ 0.007 & 
  9.761$\times10^{-11}$ $\pm$  4.728$\times10^{-12}$\\
  0.531  &  0.501  -  0.562  &  11823 & 0.169 $\pm$ 0.007 & 
  7.168$\times10^{-11}$ $\pm$  3.120$\times10^{-12}$\\
  0.596  &  0.562  -  0.631  &  8476 &	0.175 $\pm$ 0.010 & 
  4.480$\times10^{-11}$ $\pm$  2.513$\times10^{-12}$\\
  0.668  &  0.631  -  0.708  &  6480 &	0.208 $\pm$ 0.008 & 
  3.524$\times10^{-11}$ $\pm$  1.372$\times10^{-12}$\\
  0.749  &  0.708  -  0.794  &  5932 &	0.170 $\pm$ 0.009& 
  2.333$\times10^{-11}$ $\pm$  1.168$\times10^{-12}$\\
  0.840  &  0.794  -  0.891  &  4146 &	0.164 $\pm$ 0.009& 
  1.376$\times10^{-11}$ $\pm$  7.186$\times10^{-13}$\\
  0.995  &  0.891  -  1.122  &  5547 &	0.140 $\pm$ 0.010& 
  6.370$\times10^{-12}$ $\pm$  4.700$\times10^{-13}$\\
  1.253  &  1.122  -  1.413  &  2951 &	0.141 $\pm$ 0.014& 
  2.689$\times10^{-12}$ $\pm$  2.721$\times10^{-13}$\\
  1.662  &  1.413  -  1.995  &  2495 &	0.107 $\pm$ 0.013& 
  8.620$\times10^{-13}$ $\pm$  1.058$\times10^{-13}$\\
  2.347  &  1.995  -  2.818  &  892 &	0.111 $\pm$ 0.013& 
  2.233$\times10^{-13}$ $\pm$  3.015$\times10^{-14}$\\
  3.315  &  2.818  -  3.981  &  529 &	0.055 $\pm$ 0.014& 
  4.645$\times10^{-14}$ $\pm$  1.216$\times10^{-14}$\\
  4.683  &  3.981  -  5.623  &  207 &	0.167 $\pm$ 0.082& 
  3.621$\times10^{-14}$ $\pm$  1.780$\times10^{-14}$\\
\hline

\end{tabular}
\end{center}

\caption{CRE Flux with statistical errors. The total number of events in each energy bin and fraction of those events that are CRE-like from the likelihood fit are also given. }
\end{table}

\section{Conclusion\label{SecDisc}}

CRE results shown here are consistent with prior ground-based and 
space-borne measurements at similar energies.
This result represents the second ground-based,
high-statistics measurement of a break in the CRE spectrum around $\sim$1 
TeV, seen by H.E.S.S., CALET and DAMPE, but not seen by MAGIC or {\it Fermi}-
LAT. 
The precise value of this break energy is an important parameter in any 
successful model of our local CRE environment.
Based on the fit of the CRE spectrum in the previous section, two power laws with 
a break between the indices best describes the data. However, a power law with an exponential cutoff is not completely ruled out. A single power law is ruled out by the VERITAS data.

Several different sources in our local neighborhood have been speculated as the 
accelerators of electrons at TeV energies, including supernova remnants and pulsars. 
The decay or annihilation of WIMP DM has also been proposed as the dominant
source of CREs and a reason for the positron fraction rising with energy, 
\cite{Bertone2005} but it is currently not possible to discriminate dark matter 
models from other sources with the available data \cite{Grasso2009}. 
Nearby pulsars with distances 
less than 1 kpc may also be sources of relativistic electrons and positrons
\cite{COWSIK}\cite{Pohl1998}. Because of 
synchrotron and inverse-Compton energy losses, the age of TeV electrons is 
$\sim$10$^5$ years, and decreases with increasing energy. Very few of the know pulsars are capable of accelerating electrons to TeV-scale energies, namely 
Geminga, Monogem and a handful of others. Breaks in the spectrum at TeV energies are 
expected as the number of astrophysical sources capable of accelerating 
CREs to those energies decreases \cite{Grasso2009}. Refined measurements of CRE 
spectra from IACTs, including VERITAS and the upcoming CTA observatory \cite{CTA}, 
should help with understanding the number and distribution of sources 
capable of accelerating CREs to TeV-scale energies.

\begin{acknowledgments}

\section{Acknowledgments}

This research is supported by grants from the U.S. Department of Energy 
Office of Science, the U.S. National Science Foundation and the Smithsonian 
Institution, and by NSERC in Canada. We acknowledge the excellent work of the 
technical support staff at the Fred Lawrence Whipple Observatory and at the 
collaborating institutions in the construction and operation of the instrument.
Computations were made on the supercomputer Guillimin from McGill University, 
managed by Calcul Quebec and Compute Canada. The operation of this 
supercomputer is funded by the Canada Foundation for Innovation (CFI), 
NanoQuebec, RMGA and the Fonds de recherche du Quebec - Nature et 
technologies (FRQ-NT). 

\end{acknowledgments}

%\bibliographystyle{plain}
%\bibliography{CRE_PRD_v1}

\end{document}